\documentclass[journal]{IEEEtran}
\usepackage{url}
\newcommand{\reals}{{\mathrm{I\kern-.2em R}}}
\newcommand{\complex}{{\mathrm{C\kern-.6em C}}}
\newcommand{\field}{{\mathrm{I\kern-.2em F}}}
\newcommand{\expectation}{{\mathrm{I\kern-.2em E}}}

\newcommand{\dd }{{\rm d}}

\newcommand{\pp}{p^\prime}

\newcommand{\qp}{q^\prime}

\newcommand{\vc}{{\bf c}}

\newcommand{\vx }{{\bf x}}
\newcommand{\vxp }{{\bf x^\prime}}

\newcommand{\vw }{{\bf w}}
\newcommand{\vy }{{\bf y}}

\newcommand{\mzero }{{\bf 0}}

\usepackage{lineno}
\usepackage{enumitem}
\usepackage{times,amsfonts,amsmath,mathtools,epsfig,fixltx2e,amsthm,amssymb}
\usepackage{graphics,graphicx,epstopdf}
\usepackage{caption}
\usepackage{url}
\usepackage{cite}
\usepackage{multirow}
\usepackage{algorithm} 
\usepackage{algorithmic} 
\usepackage[algo2e,linesnumbered,ruled,vlined]{algorithm2e}

\modulolinenumbers[5]

\newcommand{\basispln}{I_{p;l,n}}

\newcommand{\scalarbasispln}{I_{p;l,n}}

\newcommand{\abstractfullbasiszeta}{F_\zeta}

\newcommand{\Nirrep}{N_{\rm rep}}

\newcommand{\Npl}{N_{p;l}}

\newcommand{\Radius}{x}

\ifCLASSINFOpdf
\else
\fi
\begin{document}
\title{3D Reconstruction of Heterogeneous Virus Particles with Statistical Geometric Symmetry}

\author{Nan~Xu,~\IEEEmembership{Student Member,~IEEE,}
        and~Peter~C.~Doerschuk,~\IEEEmembership{Member,~OSA,~IEEE}% <-this % stops a space
\thanks{N. Xu is with the School
of Electrical and Computer Engineering, Cornell University, Ithaca, NY, 14853 USA e-mail: nx25@cornell.edu}% <-this % stops a space
\thanks{P.C. Doershuck is with the School
	of Electrical and Computer Engineering and the Nancy E. and Peter C. Meinig School
	of Biomedical Engineering, Cornell University, Ithaca, NY, 14853 USA e-mail: pd83@cornell.edu}% <-this % stops a space
}

\markboth{Journal of IEEE Transactions on 
	Computational Imaging, for review purposes, Novomber~2016}%
{Xu \MakeLowercase{\textit{et al.}}: Symmetric Statistical Characterization of Heterogeneous Virus Particles}

\maketitle

\begin{abstract}
In 3-D reconstruction problems, the image data obtained from cryo electron microscopy is the projection of many heterogeneous instances of the object under study (e.g., a virus). When the object is heterogeneous but has an overall symmetry, it is natural to describe the object as stochastic with symmetrical statistics. This paper presents a maximum likelihood reconstruction approach which allows each object to lack symmetry while constraining the {\it statistics} of the ensemble of objects to have symmetry. This algorithm is demonstrated on bacteriophage HK97 and is contrasted with an existing algorithm in which each object, while still heterogeneous, has the symmetry. Reconstruction results show that the proposed algorithm eliminates long-standing distortions in previous heterogeneity calculations associated with symmetry axes, and provides estimates that make more biologically sense than the estimates of existing algorithms.
\end{abstract}


\IEEEpeerreviewmaketitle

\section{Introduction}
%In virology, there is a large class of plant and animal viruses called ``spherical'' viruses, 
\IEEEPARstart{I}{n} virology, discovering the structure of virus particles has been important. The virus particle has a shell of protein, called a ``capsid'', which surrounds a cavity containing the viral genome. Typical sizes and molecular weights of the virus particles are $10^2$--$10^3${\AA} and 10MDa. The capsid is constructed of many repetitions of the same peptide molecule in geometric arrays. %similar to human-constructed geodesic domes. 
The geometry of these spherical viruses is important to their lifecycles. Cryo electron microscopy (cryo EM), which has become an important technique for determining the geometry of a particle, leads to 3-D image reconstruction problems for these biological nanomachines. Specifically, $10^3-10^6$ virus particles are flash frozen to cryogeneic temperatures and imaged. Each image is basically a highly-noisy ($\mbox{SNR}<0.1$) 2-D projection of the 3-D electron scattering intensity distribution of the particle. Only one such image can be taken per particle and its projection orientation is unknown. Therefore, the information from many such images is fused to compute a 3-D reconstruction~\cite{JensenMethodsEnzymologyABC2010}.
\par
In the maturation process, the virus particles can be grouped into a number of discrete classes depending on their geometric forms, sizes, pH values, etc. In~\cite{QiuWangMatsuiDomitrovicYiliZhengDoerschukJohnsonJSB2012,YiliZhengQiuWangDoerschukJOSA2012}, a reconstruction approach was presented, which includes both discrete classes and continuous variability of the particles within each class. The existing cryo electron microscopy are able to experimentalize an ensemble of virus particles within one class, which will be the objects to be discussed in the remainder of this paper.
%The assembly of virus particles is achieved by the information contained within the components of the particle, a process driven by the rules of symmetry and by the most thermodynamically stable configuration. 
\par
In cryo-electron microscopy, the 3-D structure of each virus particle is described by the 3-D electron scattering intensity function, denoted by $\rho(\cdot)$. The capsid of a virus particle has been well known to obey certain geometric symmetry~\cite{baron1996epidemiology}. The basic and the most popular assumption of the virus particles within one class~\cite{DoerschukJohnsonIT2000,YinZhengDoerschukNatarajanJohnsonJSB2003,ScheresGaoValleHermanEggermontFrankCarazoNatureMethods2007,JLeeDoerschukJohnsonIP2007} is that each instance of the virus particle is identical and exactly obeys the rotational symmetry. In particular, 
\begin{equation}
\rho(\vx)=\rho_0(\vx) \mbox{~with~} \rho_0(\vx)=\rho_0(M^{-1}\vx),
\end{equation}
where $M$ is the rotational matrix and $\vx\in\reals^3$ are real-space coordinates. We call it assumption of ``identical individuals''. The reconstruction algorithm based on this rudimentary assumption is usually called the ``homogeneous reconstruction with identical individuals'' method (HRII). A sufficient assumption introduced in~\cite{QiuWangMatsuiDomitrovicYiliZhengDoerschukJohnsonJSB2012,YiliZhengQiuWangDoerschukJOSA2012} is that instances are different from each other (for example, due to the inherent flexibility of such a large molecular complex) but still each instance exactly obeys the rotational symmetry. In particular,
\begin{equation}
\rho(\vx)=\rho(M^{-1}\vx).
\end{equation} In the remainder of this paper, this assumption is referred as the assumption of ``non-identical symmetrical individuals'', and the reconstruction algorithm~\cite{QiuWangMatsuiDomitrovicYiliZhengDoerschukJohnsonJSB2012,YiliZhengQiuWangDoerschukJOSA2012} developed based on this assumption is named as the ``heterogeneous reconstruction with symmetrical individuals'' (HRSI).
\begin{figure}[h]
	\begin{center}
		\includegraphics[width=7cm]{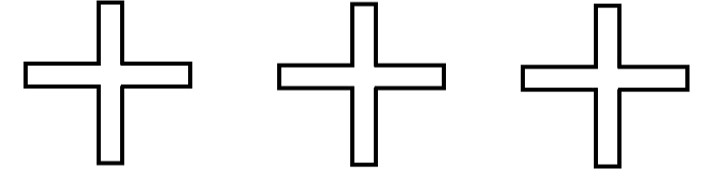}\\
		(a) Identical individuals (HRII)\\
		\includegraphics[width=7cm]{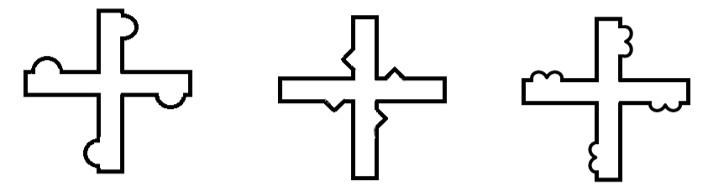}\\
		(b) Non-identical symmetrical individuals (HRSI)\\
		\includegraphics[width=7cm]{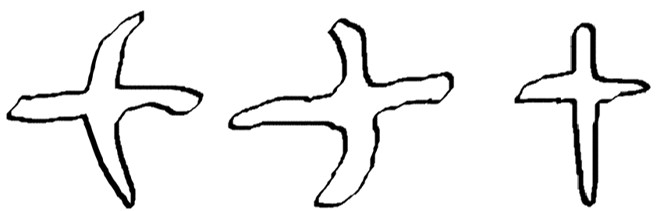}\\
		(c) Symmetrical statistics (HRSS)
	\end{center}
	\caption{\label{fig:SymParticle}
		Four fold symmetry particles based on the three different assumptions and algorithms.		
	}
\end{figure}
\par
In contrast to this previous work, this paper considers a more realistic and sophisticated view: The different instances of the particle are different and it is the {\it statistics} of the particle that obey the symmetry not the individual instances. In particular, the mean and covariance of the 3-D 
structure of the virus particle are invariant under the rotations. This idea is referred as  the ``symmetrical statistics''. 
\begin{equation}
\bar\rho(M^{-1}\vx)=\bar\rho(\vx) \mbox{~and~} C_\rho(M^{-1}\vx,M^{-1}\vxp)=C_\rho(\vx,\vxp) \label{eq:symstats}
\end{equation} where $\bar{\rho}(\vx)=\expectation[\rho(\vx)]$ and $C(\vx,\vxp)=\expectation[\rho(\vx)\rho(\vxp)]$. We call the new reconstruction algorithm as ``heterogeneous reconstruction with symmetrical statistics'' (HRSS). Three possible structures described by each of the three assumptions are demonstrated through an example of four fold symmetry in Fig.~\ref{fig:SymParticle}.
\par 
The largest class of plant and animal viruses is ``spherical'' viruses. The capsid of spherical virus particles most common is icosahedral symmetry--the symmetry group of an icosahedron--which is invariant under a total of 60 rotational symmetry operations. In 3-D reality, the structure of a virus particle can be expressed as a linear combination of basis functions where the weights in the linear combination are Gaussian random variables~\cite{QiuWangMatsuiDomitrovicYiliZhengDoerschukJohnsonJSB2012,YiliZhengQiuWangDoerschukJOSA2012}. In particular, %\vspace*{-3mm}
\begin{equation}
\rho(\vx)=\sum_{\zeta} \abstractfullbasiszeta(\vx) c_{\zeta}\vspace*{-2mm}
\label{eq:orthonormalexpansion}
\end{equation}
where $\abstractfullbasiszeta(\vx)$ is the basis function, $c_{\zeta}$ is the weight, and $\vx\in\reals^3$ are real-space coordinates. Because $\rho(\cdot)$ is real, it is convenient to have real-valued basis functions and weights. 
\par 
To describe the situations of both ``identical individuals'' and ``non-identical symmetrical individuals'', the complete set of symmetric basis functions of the icosahedral group is sufficient. In the above rudimentary example of 2-D particles with four fold symmetry, these two situations (Fig.~\ref{fig:SymParticle}(a) and Fig.~\ref{fig:SymParticle}(b)) can be both described by the basis functions $e^{i4n\theta}h_l(r)$, where $h_l(r)$ is spherical Bessel functions for degree $l\in\mathbb{N}$. However, to describe the situation in which particles have asymmetric structures but symmetric statistics, a complete orthonormal set of real-valued basis functions with specific rotational properties under the operation of the icosahedral group is required (Section \ref{sec:BasisFunction}).
\par 
The maximum likelihood (ML) reconstruction algorithm can
estimate the mean vector and covariance matrix of the vector of weights $c_{\zeta}$, which is a generalization of classical ML Gaussian mixture parameter estimation~\cite{RednerWalker1984}. After choosing appropriate basis functions, the estimates of the mean vector and covariance matrix need to be constrained in order to realize the two statistical symmetries as shown in Eq.~\ref{eq:symstats} (Section \ref{sec:covariance}). Taking the advantage of the existing reconstruction algorithms, the HRSS algorithm, introduced in this paper, can be developed by adding new basis functions and the additional constrains for the the mean vector and covariance matrix of the weights to the existing HRSI algorithm (Section \ref{sec:MLE}). The reconstruction results are demonstrated in constrast to the previous algorithms in Sections \ref{sec:results}. As a result, HRSS eliminates the long-standing distortions in heterogeneity calculations associated with symmetry axes introduced by former algorithms. The whole paper is concluded in Section \ref{sec:conc}.
\section{Real-valued basis functions}
\label{sec:BasisFunction}
Each real-valued basis function $\abstractfullbasiszeta(\vx)$ is a product of an real-valued angular basis function and a real-valued radial basis function. The real-valued angular basis functions are complete orthonormal square integrable functions on the surface of the sphere, which includes the symmetry properties of the particle. In particular, they are the complete set of real-valued basis functions of the icosahedral group, which were derived and computed in \cite{XuBasisFunctions2016} via generalized projection from group theory. Specifically, projection operators are applied to the real-valued spherical harmonics\footnote{Spherical harmonics are denoted by $Y_{l,m}(\theta,\phi)$ where the degree $l$ satisfies $l\in\mathbb{N}$, the order $m$ satisfies $m\in\{-l,\dots,l\}$ and $(\theta,\phi)$ are the angles of spherical coordinates with $0\leq \theta\leq \pi$ and $0\leq \phi\leq 2\pi$~\cite[Section~14.30, pp.~378--379]{OlverLozierBoisvertClark2010}.}~\cite[p.~93]{Cornwell1984}, which computes a basis function of the icosahedral group as a linear combination of spherical harmonics of a fixed degree $l$. 
\par
A key concept used in this approach is the irreducible representations (irreps) of a finite group, which are sets of 2-D square matrices that are homomorphic under matrix multiplication to the group elements. For the icosahedral group, it is possible to find all irreps matrices that are real-valued, which is required to generate the compete orthonormal basis functions that are real-valued \cite{XuBasisFunctions2016}. Suppose that $N_{\rm rep}$ is the number of irreps of the $N_g$-order group $G$. Let the set of real-valued matrices in the $p$th irrep be denoted by $\Gamma^p(g)\in\mathbb{R}^{d_p\times d_p}$ for all $g\in G$ where $p\in\{1,\dots,N_{\rm rep}\}$. For the icosahedral group, $N_{\rm rep}=5$, $N_g=60$, and $d_p=1,3,3,4,5$ for $p=1,2,3,4,5$, respectively.
\par
\begin{figure}[h!]
	\begin{center}
		\begin{tabular}{ccc}
			%came from ~/p318.kangwang.chiyufu.lambdaportals.dirtex/icosahedron.trim.eps
			\includegraphics[width=2.4cm]{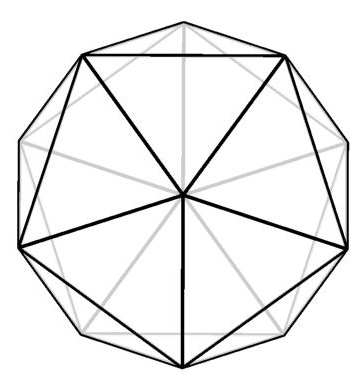}
			&
			\includegraphics[width=2.4cm]{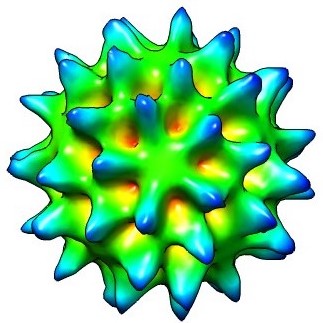}
			&
			\includegraphics[width=2.4cm]{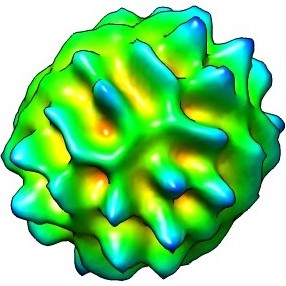}
			\\
			&
			$p=1$
			&
			$p=2$
			\\
			\includegraphics[width=2.4cm]{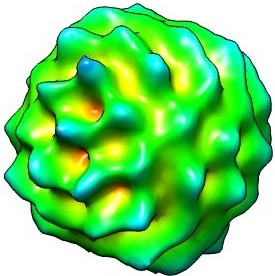}
			&
			\includegraphics[width=2.4cm]{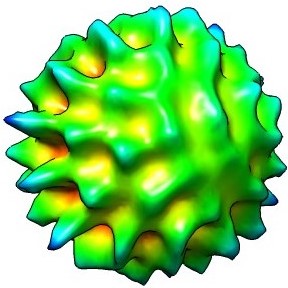}
			&
			\includegraphics[width=2.4cm]{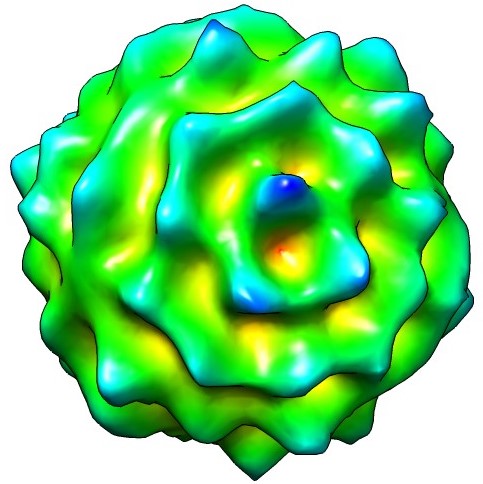}
			\\
			$p=3$
			&
			$p=4$
			&
			$p=5$
		\end{tabular}
	\end{center}
	\caption{
		\label{fig:irredrep}
		An icosahedron with one of each type of symmetry axis (2-, 3-, and 5-fold)
		shown and example angular basis functions with $l=10$ and
		$p\in\{1,\dots \Nirrep\}$.
		The surfaces of 3-D objects defined by $\xi(\vx)=1$ for $x\le \kappa_1+\kappa_2 \scalarbasispln(\vx/x)$ and 0 otherwise, where $\kappa_1$ and $\kappa_2$ are chosen so that $\kappa_1+\kappa_2
		\scalarbasispln(\vx/x)$ varies between 0.5 and 1.
		are visualized by UCSF
		%Chimera~\cite{PettersenHuangCouchGreenblattMengFerrin2004} where the color
		Chimera where the color
		indicates the distance from the center of the object.	%	\vspace*{-5mm}
	}
\end{figure}

The resulting real-valued angular basis functions associated with the degree $l$ and the $p$th irrep, denoted by $I_{p;l,n}(\vx/x)$ for $n\in\{1,\dots,N_{p;l}\}$, have the following properties.
\begin{enumerate}[itemsep=0mm]
	\item
	Each $I_{p;l,n}$ is a $d_p$-dimensional real-valued vector function, i.e.,
	$I_{p;l,n}\in \mathbb{R}^{d_p}$.
	\item
	The $I_{p;l,n}$ functions are orthonormal on the surface of the sphere;
	\item
	The subspace of square integrable functions on the surface of the sphere defined by spherical harmonics of index $l$, contains a set of
	$I_{p;l,n}$ functions with a total of $2l+1$ components.
	\item
	Each $I_{p;l,n}$ function has a specific transformation property under rotations from the icosahedral group~\cite[p.~20]{Cornwell1984}, in
	particular,
	\begin{equation}
	I_{p;l,n}(R_g^{-1} \vx/x)
	=
	(\Gamma^p(g))^T
	I_{p;l,n}(\vx/x)
	\label{eq:transformationEqn}
	\end{equation}
	where ${}^T$ is transpose not Hermitian transpose and $R_g$ is the
	$3\times 3$ rotation matrix corresponding to the $g$th element of the
	group.
\end{enumerate}
These properties were numerically verified from the close relationship between the $I_{p;l,n}$ and spherical harmonics. Furthermore, it follows that the family of $I_{p;l,n}$ is a complete basis for square integrable functions on the surface of the sphere. Examples of $I_{p,j;l,n}$ functions for $l=15$ are visualized in Figure~\ref{fig:irredrep}. Note that $p=1$ exhibits all of the symmetries of an icosahedron. The angular basis functions used in the two previous algorithms, HRII and HRSI, are just the angular basis functions at $p=1$.
%\subsection{Real-valued radial basis functions}
\par 
The radial basis functions $h_{l,q}(\Radius)$ for $q\in\{1,2,\dots\}$  are exactly the family of Bessel functions used in~\cite[Section~IIIB]{ZhengDoerschukIP1996}, which are real-valued and have the orthonormal condition, specifically, 
\begin{equation}\label{eq:radialortho}
\int_{r=0}^\infty
h_{l,q}(\Radius) h_{l,\qp}(\Radius) \Radius^2 \dd\Radius=
\delta_{q,\qp}.
\end{equation}
The reason that $h_{l,q}$ depends on $l$ is that this makes it possible to have symbolic formulas for the 3-D Fourier transform of $\rho(\vx)$.
%\par
%Peter's radial basis functions paragraph...
\par 
Let $\zeta$ be the shorthand for $l,n,q$, which has the total number that associates with the $p$th irrep of the icosahedral group to be $N_{p;\zeta}$, i.e.,  $\sum_{\zeta=1}^{N_{p;\zeta}}=\sum_{l=1}^{\infty}\sum_{n=1}^{\Npl}\sum_{q=1}^{\infty}$. Each real-valued basis function, denoted by $F_{p;\zeta}(\vx)$, is the product of the $n$th angular basis function and the $q$th radial basis function that associate with the $p$th irrep of the icosahedral group and $l$th degree of spherical harmonics, specifically, $F_{p;\zeta}(\vx)=I_{p;l,n}(\vx/x)h_{l,q}(\Radius)$. The real-valued basis functions, $F_{p;\zeta}$ for $p\in\{1,\dots,\Nirrep\}$ and $\zeta\in\{1,\dots,N_{p;\zeta}\}$ have the following properties.
\begin{enumerate}[itemsep=0mm]
	\item[i)]
	Each $F_{p;\zeta}$ is a $d_p$-dimensional real-valued vector function, i.e, $F_{p;\zeta}\in\mathbb{R}^{d_p}$.
	\item[ii)]
	The $F_{p;\zeta}$ functions construct a complete orthonormal basis because of the orthonormality of $I_{p;l,n}$ and $h_{l,q}$, specifically, 
	\begin{equation}\label{eq:Fortho}
	\int_{\vx}
	F_{p;\zeta}(\vx) F_{p';\zeta'}^T(\vx)\dd\vx=
	I_{d_p}\delta_{p,\pp}\delta_{\zeta,\zeta'}.\end{equation}
	(Please see Appendix A for the derivation.)
	\item[iii)]
	Each $F_{p;\zeta}$ satisfies the following transformation because of Eq.~\ref{eq:transformationEqn}, specifically, for all $g\in G$,
	\begin{eqnarray}
	F_{p;\zeta}(R_g^{-1}\vx)%&=&F_{p;l,n,q}(R_g^{-1}\vx)\nonumber\\
	&=&\basispln(R_g^{-1} \vx/x)h_{l,q}(\Radius)\nonumber\\
	&=&
	(\Gamma^p(g))^T
	\basispln(\vx/x)h_{l,q}(\Radius)\nonumber\\
	%	&=&(\Gamma^p(g))^T	F_{p;l,n,q}(\vx)\nonumber\\
	&=&(\Gamma^p(g))^T	F_{p;\zeta}(\vx).
	\label{eq:FtransEqn}
	\end{eqnarray}
\end{enumerate}
\iffalse (1) it is a $d_p$-dimensional real-valued vector function, i.e, $F_{p;\zeta}(\vx)\in\mathbb{R}^{d_p}$, and (2) it constructs a complete orthonormal basis because of the orthonormality of $I_{p;l,n}$ and $h_{l,q}$, specifically, 
\begin{equation}\label{eq:Fortho}
\int_{\vx}
F_{p;l,n,q}(\vx) F_{p';l',n',q'}(\vx)\dd\vx=
I_{d_p}\delta_{p,\pp}\delta_{l,l'}\delta_{n,n'}\delta_{q,\qp},
\end{equation}
and (3) it satisfies the following transformation because of Eq.~\ref{eq:transformationEqn}, specifically, for all $g\in G$,
\begin{eqnarray}
F_{p;\zeta}(R_g^{-1}\vx)%&=&F_{p;l,n,q}(R_g^{-1}\vx)\nonumber\\
&=&\basispln(R_g^{-1} \vx/x)h_{l,q}(\Radius)\nonumber\\
&=&
(\Gamma^p(g))^T
\basispln(\vx/x)h_{l,q}(\Radius)\nonumber\\
%	&=&(\Gamma^p(g))^T	F_{p;l,n,q}(\vx)\nonumber\\
&=&(\Gamma^p(g))^T	F_{p;\zeta}(\vx).
\label{eq:FtransEqn}
\end{eqnarray}
\fi
\section{Constraints on weights caused by symmetrical statistics} 
\label{sec:covariance}
The weight corresponding to the basis function $F_{p;\zeta}$, denoted by $c_{p;\zeta}$, is a real-valued vector, i.e., $c_{p;\zeta}\in\mathbb{R}^{d_p}$.  Eq.~\ref{eq:orthonormalexpansion} can be restated as
{\small
	\begin{eqnarray}
	%\begin{split}
	\rho(\vx)
	&=&
	\sum_{p=1}^{\Nirrep}\sum_{\zeta=1}^{N_{p;\zeta}} F^T_{p;\zeta}(\vx)c_{p;\zeta}.\label{eq:orthonormalexpansion2a}
	\end{eqnarray}
}
The index $\tau$ in Eq.~\ref{eq:orthonormalexpansion} is the shorthand of $p$ and $\zeta$, which has the total number $N_\tau$ to satisfy $N_\tau=\sum_{p=1}^{\Nirrep}N_{p;\zeta}$. The vector of weights $\vc=[c_1^T,\dots,c_{N_\tau}^T]^T=[c_{N_{p=1,\zeta}}^T, ..., c_{N_{p=1,\zeta}}^T,\dots,c_{N_{p=\Nirrep,1}}^T,... c_{N_{p=\Nirrep,\zeta}}^T]^T$ has Gaussian distribution $N(\bar{\vc}, V)$, where $\bar{\vc}=\expectation[\vc]\in\mathbb{R}^{N_\tau\times 1}$ and $V=\expectation[\vc\vc^T]\in\mathbb{R}^{N_\tau\times N_\tau}$. 

Our assumption of symmetrical statistics for the spherical virus particles is that the mean and the correlation function of the 3-D scattering intensity are invariant under all 60 rotations of the icosahedral group, which can be translated into the following two conditions, for all $g\in G$,
\begin{eqnarray}
\bar\rho(R_g^{-1}\vx)&=&\bar\rho(\vx), \label{eq:symmean}\\
C_\rho(R_g^{-1}\vx,R_g^{-1}\vxp)&=&C_\rho(\vx,\vxp), \label{eq:symcov}
\end{eqnarray}
where $\bar{\rho}(\vx)=\expectation[\rho(\vx)]$, $C(\vx,\vxp)=\expectation[\rho(\vx)\rho(\vxp)]$ and $R_g$ is the $g$th rotation matrix in group $G$. These two conditions induce constraints on the weights of basis functions, in particular, the mean vector $\bar{\vc}$ and covariance matrix $V$.
\subsection{Constraint caused by the first order statistics}
The symmetry on the first order statistics (Eq.~\ref{eq:symmean}) induces a constraint on the mean of the weights.  By definition (Eq. \ref{eq:orthonormalexpansion2a}) and the transformation property (Eq.~\ref{eq:FtransEqn}), the RHS of Eq.~\ref{eq:symmean} can be expanded as
{\small
	\begin{equation}
	\label{eq:RhobarR}
	\begin{split}
	\bar{\rho}(\vx)&=\expectation\left[
	\sum_{p}\sum_{\zeta} F^T_{p;\zeta}(\vx)
	c_{p;\zeta}\right]\\
	&=\sum_{p}\sum_{\zeta} F^T_{p;\zeta}(\vx)
	\expectation\left[c_{p;\zeta}\right]
	\end{split}
	\end{equation}
}

and the LHS of Eq.~\ref{eq:symmean} becomes
{\small
	\begin{equation}
	\begin{split}
	\bar{\rho}(R_g^{-1}\vx)&=\sum_{p}\sum_{\zeta} F_{p;\zeta}^T(R_g^{-1}\vx)\expectation\left[c_{p;\zeta}\right]\\
	&=\sum_{p}\sum_{\zeta} \left(\Gamma^p(g)^TF_{p;\zeta}(\vx)\right)^T
	\expectation\left[c_{p;\zeta}\right]\\
	&=\sum_{p}\sum_{\zeta} F^T_{p;\zeta}(\vx)
	\Gamma_c^p(g)
	\expectation\left[c_{p;\zeta}\right],
	\end{split}
	\label{eq:RhobarL}
	\end{equation}
}

Equating Eq.~\ref{eq:RhobarL} and Eq. \ref{eq:RhobarR} to give
{\small
	\begin{equation}
	\sum_{p}\sum_{\zeta} F^T_{p;\zeta}(\vx)
	(\Gamma_c^p(g)-I_d)
	\expectation\left[c_{p;\zeta}\right]=0,
	\end{equation}
} which holds for all $g\in G$. Because only the $\Gamma_c^{p=1}(g)=1$ for all $g\in G$ and $\Gamma_c^{p\neq1}(g)$ for all $g\in G$ are real-valued full matrices, the mean of $c_{p;\zeta}$, needs to satisfy
\begin{equation}
\expectation\left[c_{p;\zeta}\right]=	\expectation\left[c_{p;\zeta}\right]\delta_{p,1}.
\label{eq:cbarconstraint:generalform}
\end{equation}
In other words, the entries of $\bar{\vc}$ that corresponding to the $p$th ($p\neq 1$) irrep are constrained to be zero.

\subsection{Constraint caused by the second order statistics}
The symmetry condition on the second order statistics (Eq.~\ref{eq:symcov})
can be achieved by constraints on the covariance matrix of the weights.

Similar to the derivation of Eq.~\ref{eq:RhobarR}-\ref{eq:RhobarL}, the RHS of Eq.~\ref{eq:symcov} can be expressed as
{\scriptsize
	\begin{equation}\label{eq:R}
	\begin{split}
	&C_\rho(\vx,\vxp)\\
	&=\expectation\left[
	\left(\sum_{p}\sum_{\zeta}F_{p;\zeta}^T(\vx)c_{p;\zeta}\right)
	\left(\sum_{p'}\sum_{\zeta'}F_{p';\zeta'}^T(\vx')c_{p';\zeta'}\right)^T
	\right]\\
	&=\sum_{p}\sum_{\zeta}\sum_{p'}\sum_{\zeta'} F_{p;\zeta}^T(\vx)
	\expectation\left[c_{p;\zeta}c_{p\;\zeta'}^T\right]
	F_{p';\zeta'}(\vx'),
	\end{split}
	\end{equation}
}
and the LHS of Eq.~\ref{eq:symcov} can be expressed as
{\scriptsize
	\begin{equation}
	\begin{split}
	& C_\rho(R_g^{-1}\vx, R_g^{-1}\vxp)\\
	&{=\sum_{p}\sum_{\zeta}\sum_{p'}\sum_{\zeta'} F_{p;\zeta}(R_g^{-1}\vx)
		\expectation\left[c_{p;\zeta}c_{p';\zeta'}^T\right]
		F_{p';\zeta'}(R_g^{-1}\vx)}\\
	&{=\sum_{p}\sum_{\zeta}\sum_{p'}\sum_{\zeta'} \left((\Gamma^p_g(g))^TF_{p;\zeta}(\vx)\right)^T
		\expectation\left[c_{p;\zeta}c_{p';\zeta'}^T\right]
		\left((\Gamma^{p'}_g(g))^TF_{p';\zeta'}(\vx')\right)}\\
	&=\sum_{p}\sum_{\zeta}\sum_{p'}\sum_{\zeta'} F^T_{p;\zeta}(\vx)\Gamma^p_g(g)
	\expectation\left[c_{p;\zeta}c_{\pp;\zeta'}^T\right]
	(\Gamma^{p'}_g(g))^TF_{p';\zeta'}(\vx')\\
	\end{split}
	\label{eq:R:rotated}
	\end{equation}
}
On both sides of the symmetry condition (Eq. \ref{eq:symcov}), multiply on the left by $F_{p;\zeta}(\vx)$ and on the right by $F_{p';\zeta'}^T(\vx')$ for random $(p,\zeta)$ and $(p',\zeta')$, and integrate over
the 3D coordinates $\vx$ and $\vxp$. Specifically,
{\small
\begin{equation}\label{eq:rxxErRxRx}
\begin{split}
&\int_{\vx}
\int_{\vxp}
F_{p;\zeta}(\vx)C_\rho(\vx,\vxp)F_{p';\zeta'}(\vx')
\dd\vx
\dd\vxp\\
&=\int_{\vx}
\int_{\vxp}
F_{p;\zeta}(\vx)C_\rho(R_g^{-1}\vx,R_g^{-1}\vxp)F_{p';\zeta'}(\vx')
\dd\vx
\dd\vxp.
\end{split}
\end{equation}
}
\par 
Because the basis functions are orthonormal on the 3D space (Eq.~\ref{eq:Fortho}), by inserting Eq. \ref{eq:R} and \ref{eq:R:rotated} into Eq. \ref{eq:rxxErRxRx}, it gives
$\expectation
\left[
c_{p;\zeta}
c_{p';\zeta'}^T
\right]
=
\Gamma^{p}(g)^T
\expectation
\left[
c_{p;\zeta}
c_{p';\zeta'}^T
\right]
\Gamma^{p'}(g)$ for all $g\in G$. Therefore, since the irred rep is orthogonal
($\Gamma^p(g)^{-1}=\Gamma^p(g)^T$), it follows that
{\small
\begin{equation}
{\Gamma^{p}(g)}\expectation
\left[
c_{p;\zeta}
c_{p';\zeta'}^T
\right]
=\expectation
\left[
c_{p;\zeta}
c_{p';\zeta'}^T
\right]
\Gamma^{p'}(g)
\label{eq:symmetry:3}
\end{equation}
}
for all
$g\in G$,
$\vx,\vxp\in \mathbb{R}^{3\times 3}$,
$p,p'\in\{1,\dots,\Nirrep\}$,
%$l_1,l_2\in\{0,1,\dots\}$,
%$n_1\in\{1,\dots,\Nplone\}$,
$\zeta\in\{1,\dots,N_{p;\zeta}\}$,
and
$\zeta'\in\{1,\dots,N_{p';\zeta'}\}$. Since $\Gamma^p(g)\in\mathbb{R}^{d_p\times d_p}$ and $\Gamma^{p'}(g)\in\mathbb{R}^{d_{p'}\times d_{p'}}$ $\expectation [c_{p,\zeta}c_{p',\zeta'}^T]$ is a $d_{p}\times d_{p'}$ matrix. Eq.~\ref{eq:symmetry:3} has substantial structure because
$\Gamma^{p}(g)$ and $\Gamma^{p'}(g)$ both for all $g\in G$ are irrep matrices.
Schur's Lemma~\cite[Theorem~I and~II, Section~4.5, p.~80]{Cornwell1984}
% and related results ~\cite[Theorem~3.4, Section~3.3, p.~69]{Miller1972}\cite[Theorem~9.6, Section~9.9, p.~326]{Artin1991}
implies that
{\small
\begin{equation}\label{eq:EmxSoln}
\expectation [c_{p,\zeta}c_{p',\zeta'}^T]=
\left\{
\begin{array}{ll}
v_{p}((\zeta, \zeta') I_{d_{p}} , & p=p' \\
\mzero_{d_{p},d_{p'}} , & \mbox{otherwise}
\end{array}\right.
\end{equation}
}
for some number $v_{p}((\zeta, \zeta')$ depending on $p$, $\zeta$ and $\zeta'$, where $\mzero_{i,j}$ is the $i\times j$ zero matrix. Note that the solution of Eq.~\ref{eq:symmetry:3} for $\expectation [c_{p,\zeta}c_{p',\zeta'}^T]$ does not depend on $\zeta$ and $\zeta'$ except that unspecified degrees of freedom in the solution could depend on $\zeta$ and $\zeta'$. 
\par 
Suppose that the indices vary from the slowest to the fastest in the order $p$, $\zeta$, or more specifically, in the order $p$, $l$, $n$, and $q$. Then $p$ is constant over sequential sets of $d_p N_{p;\zeta}$ rows and columns in covariance matrix $V$. The matrix $\expectation [c_{p,\zeta}c_{p',\zeta'}^T]$ constructs the $(p, p')^{th}$ block of the covariance matrix $V$, denoted by  $V_{p:p'}$. Precisely, $\expectation [c_{p,\zeta}c_{p',\zeta'}^T]$ is the $(\zeta,\zeta')$th sub-block of the matrix $V_{p:p'}$. Therefore, implied by Eq.~\ref{eq:EmxSoln}, the covariance matrix $V$ is a diagonal block matrix with five blocks, corresponding to the five values of $p$, and off diagonal blocks are zeros, i.e., $V_{p:p'\neq p}=\mzero_{(d_pN_{p;\zeta}),(d_{p'}N_{p';\zeta'})}$. The $p$th block itself ($V_{p:p}$) is a Kronecker product of the identity matrix $I_{d_p}$ and a $N_{p;\zeta}\times N_{p;\zeta}$ matrix $V_p$, which has the $(\zeta,\zeta')$th ($\zeta,\zeta'\in\{1,\dots,N_{p;\zeta}\}$) entry to be $v_p(\zeta,\zeta')$. More explicitly, 
{\scriptsize
\begin{equation}
\begin{split}
& V_{p:p}\\
&=V_p\otimes I_{d_p}\\
&=\left[
\begin{array}{cccc}
v_p(1,1)
&
v_p(1,2)
&
\dots
&
v_p(1,N_{p;\zeta})
\\
v_p(2,1)
&
v_p(2,2)
&
\dots
&
v_p(2,N_{p;\zeta}) \\
\vdots & \vdots & \ddots & \vdots \\
v_p(N_{p;\zeta},1)
&
v_p(N_{p;\zeta},2)
&
\dots
&
v_p(N_{p;\zeta},N_{p;\zeta})
\end{array}
\right]\otimes  I_{d_p}\\
&=\left[
\begin{array}{cccc}
v_p(1,1) I_{d_p}
&
v_p(1,2) I_{d_p}
&
\dots
&
v_p(1,N_{p;\zeta}) I_{d_p}
\\
v_p(2,1) I_{d_p}
&
v_p(2,2) I_{d_p}
&
\dots
&
v_p(2,N_{p;\zeta})  I_{d_p}\\
\vdots & \vdots & \ddots & \vdots \\
v_p(N_{p;\zeta},1) I_{d_p}
&
v_p(N_{p;\zeta},2) I_{d_p}
&
\dots
&
v_p(N_{p;\zeta},N_{p;\zeta}) I_{d_p}
\end{array}
\right].
\end{split}
\label{eq:V:oneblock}
\end{equation}
}
\par
The covariance matrix $V$ constructed from all of the $V_{p:p}$ blocks must be a real-valued positive semidefinite matrix, which requires each of the blocks is a real-valued positive semidefinite matrix. In order to make Eq.~\ref{eq:V:oneblock} be real-valued and positive semidefinite, it is necessary that $v_p(\zeta,\zeta')\in\mathbb{R}^{+}\cup \{0\}$ for all $\zeta,\zeta'\in\{1,\dots,\Nirrep\}$ and $p\in\{1,\dots,\Nirrep\}$.
%It may be superior to have slowest to fastest in the order $p$, $j$, $l$,
%$n$, and $q$.
%\par
%for
%$i\in\{1,...,N_{\zeta,q}\}$ and $p\in\{1,...,\Nirrep\}$. In addition, the off diagonal terms must decrease sufficiently quickly.

\section{Heterogeneous reconstruction with symmetrical statistics (HRSS)}
\label{sec:MLE}
The 3D reconstruction process of heterogeneous virus particles which realize the symmetrical statistics is described in this section. In particular, the first two order statistics of the weights $\bar{\vc}$ and $V$ are solved by a constrained maximum likelihood estimator.

Suppose that a image stack includes $N_y$ numbers of 2D projection images. Through the projection slice theorem, the model for the 3D structure of the virus particle (Eq.~\ref{eq:orthonormalexpansion}) can be linearly transformed into a model for the 2D projection images, specifically~\cite{YiliZhengQiuWangDoerschukJOSA2012}, 
\begin{equation}
\vy_i=L(z_i)\vc_i+\vw_i, \vw_i\sim N(0,Q)
\label{eq:2Dmodel}
\end{equation}
where $\vy_i$ is the $i$th 2D projection image in the reciprocal space, 
$z_i$ is a vector of nuisance parameters which include the projection orientation of the $i$th image and the projection location of the particle in the $i$th image, $\vc_i\in\mathbb{R}^{N_\tau}$ is the vector of weights of the $i$th image\footnote{The weights $\vc_i$ in Eq. \ref{eq:2Dmodel} are identical to the weights in the 3D model (Eq.~\ref{eq:orthonormalexpansion}).}, $L(z_i)$ is the transformation matrix from weights to the $i$th reciprocal-space image which is transformed from the real-valued basis functions by projection slice theorem, and $\vw_i$ is the white Gaussian noise of the $i$th image which has zero mean and a covariance $Q$. Based on our assumption, the vector of weights is Gaussian distributed, specifically, $\vc_i\sim N(\bar{\vc},V)$, where $\bar{\vc}$ and $V$ are the parameters to be determined.
\par
We use a maximum likelihood estimator, as is described in ~\cite{YiliZhengQiuWangDoerschukJOSA2012}, to determine $\bar{\vc}$ and $V$. The maximum likelihood estimator is computed by an expectation-maximization
algorithm with the nuisance parameters $\{z_i\}_{i=1}^{N_y}$.
Three quantities are updated, the {\em a~priori} distribution of the ensemble of virus particle, the mean $\bar{\vc}$ and the covariance $V$ of the weights. In our problem, the two quantities, $\bar{\vc}$ and $V$, are updated alternatively in the expectation-maximization algorithm. 
\par
The previous algorithm HRSI~\cite{QiuWangMatsuiDomitrovicYiliZhengDoerschukJohnsonJSB2012,TangKearneyQiuWangDoerschukBakerJohnsonJMolRecog2014}, which only used a subset of the proposed angular basis functions (just $p=1$),
used Matlab's {\tt fmincon} (option ``trust-region-reflective'') with symbolic cost, gradient of the cost, and Hessian of the cost. We desired to modify the software to include the full set of basis functions described in Section \ref{sec:BasisFunction}. The constraint on the mean vector (Eq. \ref{eq:cbarconstraint:generalform}) requires its entries to be zero if $p\neq 1$. The constraints on the covariance matrix $V$  (Eq.~\ref{eq:V:oneblock}) require certain matrix elements to be equal. To implement the above constraints, the method of computing the gradient and Hessian of the cost is modified based on the chain rule: if $f$ is the cost then 
{\small
\begin{eqnarray}
\hspace*{-3mm}\frac{
	\partial f
}{
\partial
\zeta_{p;l,n,q}
}=
\sum_{j=1}^{d_p}
\frac{
	\partial f
}{
\partial
\nu_{p,j;l,n,q}\nonumber
}, 
\label{eq:gradWRTzeta}
\end{eqnarray}
\begin{eqnarray}
\frac{
	\partial^2 f
}{
\partial
\zeta_{p_1;l_1,n_1,q_1}
\partial
\zeta_{p_2;l_2,n_2,q_2}
}
%\hspace*{-.2in}=
=
\sum_{j_1=1}^{d_{p_1}}
\sum_{j_2=1}^{d_{p_2}}
\frac{
	\partial^2 f
}{
\partial
\nu_{p_1,j_1;l_1,n_1,q_1}
\partial
\nu_{p_2,j_2;l_2,n_2,q_2}\nonumber
}
\label{eq:HWRTzeta}
.
\end{eqnarray}
}
These equations compute the necessary gradient and Hessian in terms of larger vectors and matrices which are then reduced in size. While this approach fits the software easily, more efficient approaches may be possible.

\section{Reconstruction Results} \label{sec:results}
The performance of the proposed algorithm, HRSS, is evaluated in this section. The reconstruction process has been performed both on a simulated data (Section \ref{sec:results:sim}) and on the experimental data of a Virus Like Particle
(VLP) derived from bacteriophage HK97 Prohead I$\textsuperscript{+pro}$ (Section \ref{sec:results:exp}). The VLP is essentially the bacteriophage minus the bacteriophage's tail leaving only the icosahedrally symmetric capsid.
The average outer radius of the capsid is 254 {$\AA$}. 
\subsection{Data processing}\label{sec:DataProcessing}
Two sets of 1200 cryo-EM images were separately processed. Each image containing one bacteriophage HK97 Prohead I$\textsuperscript{+pro}$ particle, was randomly selected from a larger stack. Each 2-D cryo-EM image measuring $200\times 200$ pixels with a pixel size of 2.76{$\AA$}. Due to limitations of our computer hardware (computational speed and memory size), it is not practical to process the entire stack. The image selection algorithm~\cite{QiuWangMatsuiDomitrovicYiliZhengDoerschukJohnsonJSB2012} guarantees on non-overlapping sets of images. For each data set, following the preprocessing procedure in~\cite{QiuWangMatsuiDomitrovicYiliZhengDoerschukJohnsonJSB2012,YiliZhengQiuWangDoerschukJOSA2012}, the 1200 cryo-EM images were transformed into 1200 2-D images in real and reciprocal space, and the reciprocal space images are the input to reconstruction algorithms.
\par
The reconstruction is achieved in the following two steps.
\begin{enumerate}[itemsep=0mm]
	\item[(a)]
	The mean vector of the weights, denoted by $\bar{\vc}_0$, is first computed by the {\sl homogeneous reconstruction} algorithm (HRII) described in~\cite{YiliZhengQiuWangDoerschukJOSA2012} which has $V_0=0$.
	\item[(b)]
	These results $\bar{\vc}_0$ and $V_0$ are used as the initial condition for the {\sl heterogeneous reconstruction} (HRSI or HRSS) in the second step to compute the final results for $\bar{\vc}$ and $V$.
\end{enumerate}

Both HRSI and HRSS are tested. When testing on the existing heterogeneous reconstruction model (HRSI), due to the computing limitation, basis functions with $l$ up to $55$ and $q$ up to $20$ were employed in both Step (a) (HRII) and Step (b) (HRSI). The number of parameters to be estimated in Step (b) for either $\bar \vc$ or $V$ is 1060. When testing on the proposed heterogeneous reconstruction model (HRSS) which use the complete basis functions, we aim to achieve a similar scale of parameters to be estimated. Basis functions with $l$ up to $10$ and $q$ up to $20$ were employed in both Step (a) (HRII) and Step (b) (HRSS). The number of parameters to be estimated in Step (b) for $\bar \vc$ or $V$ is 2020. 

\subsection{Performances on the simulated data} \label{sec:results:sim}
\begin{figure}[t]
	\centering\includegraphics[height=70mm]{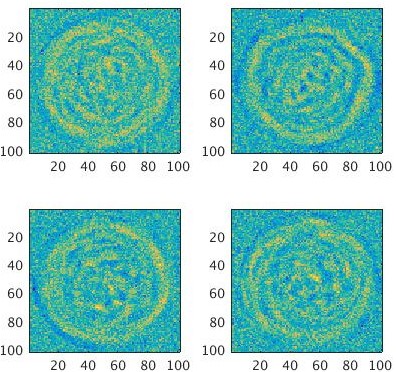}
	\caption{\label{fig:simImages} Four simulated real space 2D images of HK97.}
\end{figure}

The simulated 2D projection images in reciprocal space are generated through the same reconstruction forward model (Eq.\ref{eq:2Dmodel}). The input of the simulator is the HRSS reconstruction results $\bar{c}$ and $V$, which are called the ``ground truth'' in the stimulation study. The parameters to be determined ahead of the simulation process include the resolution (e.g., the pixel size in $\AA$ and the total number of pixels) and the signal-to-noise ratio (SNR) of the image. The method of generating the simulated 2D images is described in Algorithm~\ref{algo:SimulationStep}.
\begin{algorithm}
	\SetAlgoLined
	\KwData{The HRSS results $\bar \vc$ and $V$}
	\KwResult{a stack of 2D projection simulated images $\{y_i\}_{i=1}^N$}
	\For{$i=1$; $i<=N$; $i++$}{
		1. generate $\vc_i\sim N(\bar \vc, V)$ and pick $z_i$
		
		2. compute the 2D image:\\
		(reciprocal space) $\vy_i=L(z_i)\vc_i$\\
		(real space) $\vy_i\rightarrow Real(\vy_i)$~\cite{YiliZhengQiuWangDoerschukJOSA2012}
		
		3. construct the white noise:\\
		compute sample variances of $Real(\vy_i)$, denoted by $s^2$\\
		generate $w_i\sim N(0, Q)$, where $Q=\frac{s^2}{SNR}I$
		
		4. generate simulated images (Eq.\ref{eq:2Dmodel}): \\
		(reciprocal space)$\vy_i=L(z_i)\vc_i+w_i$\\
		(real space) $\vy_i\rightarrow Real(\vy_i)$~\cite{YiliZhengQiuWangDoerschukJOSA2012}
	}
	\caption{\label{algo:SimulationStep} Generation of $N$ simulated 2D images.}
\end{algorithm}
Let the image have $100\times 100$ pixels with a pixel size of 5.52 {$\AA$} and let $SNR=0.5$. A stack of 1200 simulated 2D projection images of HK97 is generated. Examples of the simulated real space 2D images of HK97 are shown in Fig.~\ref{fig:simImages}. With this input, the two-step reconstruction with the use of HRSS is performed. By using the same parameter settings in the reconstruction with the experimental images as the input, basis functions with $l$ up to $10$ and $q$ up to $20$ were employed in both Step (a) (HRII) and Step (b) (HRSS). 

\par
\begin{figure}
	\centering
	\begin{tabular}{cc}
		%came from ~/p318.kangwang.chiyufu.lambdaportals.dirtex/icosahedron.trim.eps
		\includegraphics[height=28mm]{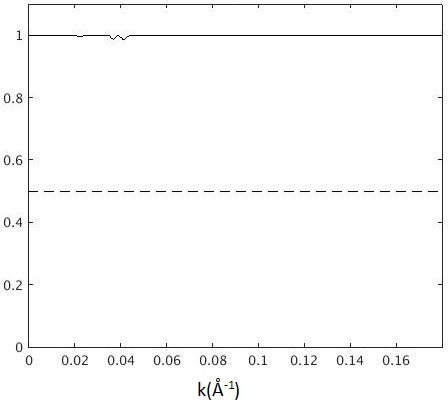}
		&%\hspace*{-1cm}
		\includegraphics[height=28mm]{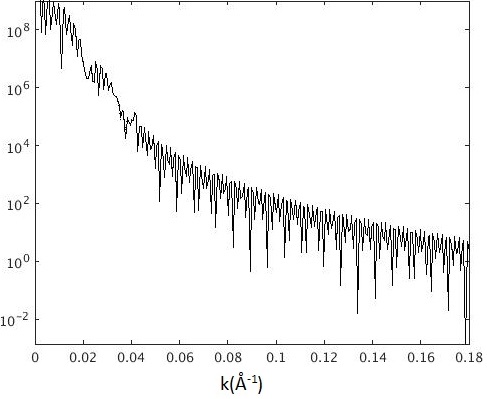}\\
		(a) FSC curve
		&%\hspace*{-1cm}
		(b) Energy curve
	\end{tabular}
	\caption{\label{fig:simFSC} FSC and energy curves between the ``ground truth'' and the reconstruction results of the simulated data.}
\end{figure}

The standard measure of performance in structural biology is
the Fourier shell correlation (FSC) (\cite{vanHeelUltramicroscopy1987}, Eq. 2; \cite{HarauzvanHeelOptik1986}, Eq. 17;
\cite{BakerOlsonFullerMicrobiolMolBiolRev1999}, p. 879) between two structures (the two estimates of mean electron scattering intensity) computed from the disjoint stacks of images. The final reconstruction ($\bar{\vc}$ and $V$) is compared with the ``ground truth''. Their FSC curve and energy curve (the denominators of FSC calculation) are shown in Fig.~\ref{fig:simFSC}. The results show that the Fourier Shell Correlation (FSC) between the reconstruction result and the ``ground truth'' are almost 1 for more than the range $k\leq0.186 \AA^{-1}$ and therefore the structures are essentially identical for more than the range $k\leq0.186 \AA^{-1}$. The energy in both structures drops around 5 at $0.186 \AA^{-1}$, which is greater than 0. Therefore, the FSC analysis in the range $k\leq0.186 \AA^{-1}$ is meaningful.

\subsection{Performances on the experimental data} \label{sec:results:exp}
\par
\begin{figure}
	\begin{center}
		\begin{tabular}{cc}
			%came from ~/p318.kangwang.chiyufu.lambdaportals.dirtex/icosahedron.trim.eps
			\hspace*{-.8cm}\includegraphics[height=28mm]{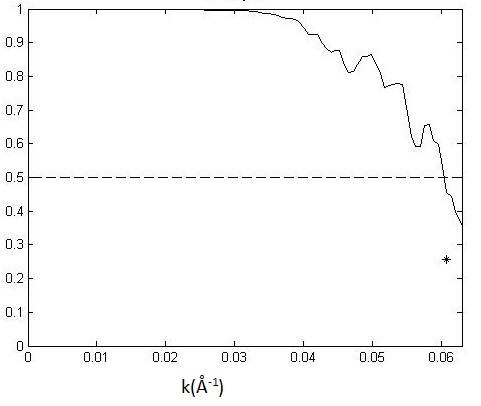}
			&%\hspace*{-1cm}
			\includegraphics[height=28mm]{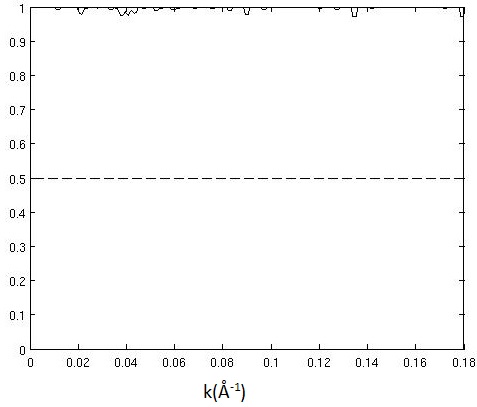}\\
			(a) symmetrical individuals
			&%\hspace*{-1cm}
			(b) symmetrical statistics \\
			(HRSI) &(HRSS)
		\end{tabular}
	\end{center}
	%\vspace*{-.15in}
	\caption{
		\label{fig:FSC}
		FSC curves between the reconstructions of the two experimental data sets for HRSI (a) and HRSS (b).
	}
\end{figure}

From the mean vector $\bar{\vc}$ and the covariance matrix $V$, it is
possible to compute the mean
function~\cite[Eq.~16]{YiliZhengQiuWangDoerschukJOSA2012},
$\bar\rho(\vx)=E[\rho(\vx)]$, and covariance
function~\cite[Eq.~18]{YiliZhengQiuWangDoerschukJOSA2012},
$C_\rho(\vx_1,\vx_2)=E[(\rho(\vx_1)-\bar\rho(\vx_1))(\rho(\vx_2)-\bar\rho(\vx_2))]$,
of the electron scattering intensity $\rho(\vx)$ of the particle.
We often visualize $s_\rho(\vx)=\sqrt{C_\rho(\vx,\vx)}$ which has the same
units as $\bar\rho(\vx)$. Through Eq.~\ref{eq:orthonormalexpansion2a}, the mean vectors $\bar \vc$ computed by HRSI and HRSS in Step (b) were used to compute two mean functions $\bar\rho(\vx)$. Similarly, the two covariance matrices $V$ computed by HRSI and HRSS were used to compute two standard deviation functions $s_\rho(\vx)$. %In addition, n high resolution homogeneous structure $\bar\rho_0(\vx)$, was computed from the mean vector, denoted by $\bar{\vc}_0$ in Step (a) with the use of basis functions with $l$ up to 55 and $q$ up do 20.
We computed FSC between the reconstructions of the two data sets for both HRSI and HRSS, which are shown in Fig.~\ref{fig:FSC}. In the case of HRSI (Fig.~\ref{fig:FSC} (a)), the resolution is determined by the fact that the FSC curve crosses the 0.5 level at 0.061 $\AA^{-1}$, which gives a resolution of 16.442 $\AA$. In the case of HRSS (Fig.~\ref{fig:FSC} (b)), the FSC between the two reconstructions are almost 1 for more than the range $k\leq0.186\AA$ and therefore the structures are essentially identical.
\begin{figure}
	\begin{center}
		\begin{tabular}{ccc}
			\includegraphics[height=22mm]{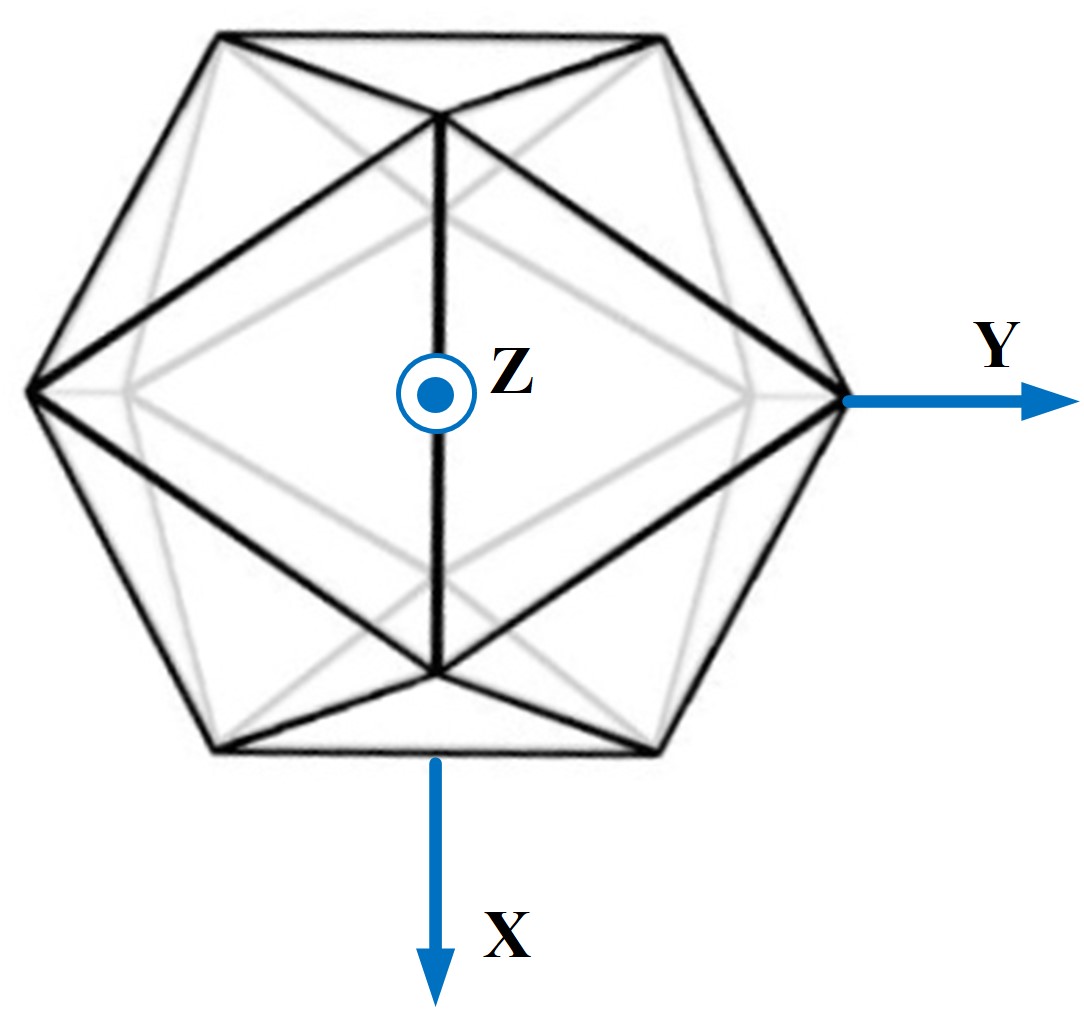}	
			&\includegraphics[height=28mm]{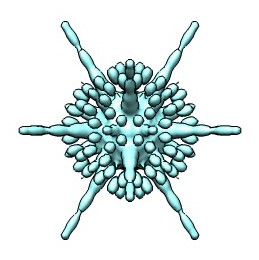}
			&\includegraphics[height=28mm]{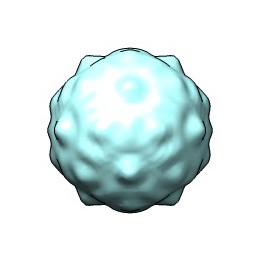}
			\\
			(a) two fold  
			&
			(b) symmetrical
			&%\hspace*{-1cm}
			(c) symmetrical\\
			symmetry axes&individuals (HRSI)&statistics (HRSS)
		\end{tabular}
	\end{center}
	%\vspace*{-.15in}
	\caption{
		\label{fig:stdv_d}
		3-D reconstructions of the standard deviation $s_{\rho}$ for HK97 Prohead
		I$\textsuperscript{Pro+}$.
		The shape is a surface of constant intensity ($0.0038$) of the standard deviation
		$s_\rho(\vx)$, which is visualized by UCSF
		Chimera~\cite{PettersenHuangCouchGreenblattMengFerrin2004}.
	}
\end{figure}

\begin{figure}[h!]
	\begin{center}
		\begin{tabular}{cc}
			%came from ~/p318.kangwang.chiyufu.lambdaportals.dirtex/icosahedron.trim.eps
			\hspace*{-.8cm}\includegraphics[width=3.2cm]{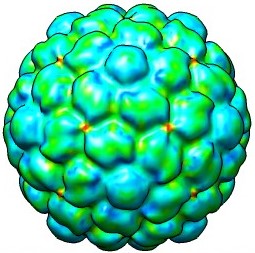}
			&%\hspace*{-1cm}
			\includegraphics[width=3.2cm]{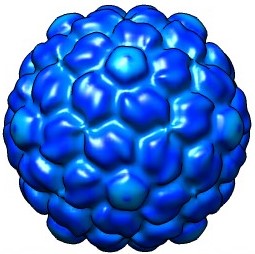}
			\\\hspace*{-.8cm}
			\includegraphics[width=4cm]{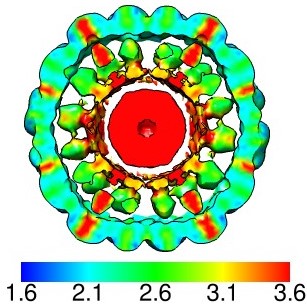}
			&%\hspace*{-1cm}
			\includegraphics[width=3.8cm]{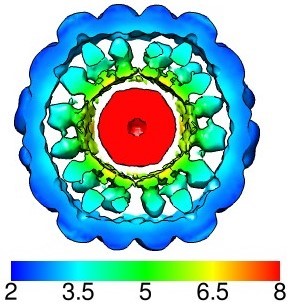}
			\\\hspace*{-.8cm}
			(a) symmetrical individuals
			&%\hspace*{-1cm}
			(b) symmetrical statistics\\
			(HRSI) &(HRSS)
			
		\end{tabular}
	\end{center}
	%\vspace*{-.15in}
	\caption{
		\label{fig:rhobar_d}
		3-D reconstructions of $\bar{\rho}$ for HK97 Prohead
		I$\textsuperscript{+pro}$ (different colormap).
		The shape is a surface of constant intensity ($5\times 10^{-4}$) of
		$\bar\rho(\vx)$ colored by
		the standard deviation $s_\rho(\vx)$, which is visualized by UCSF
		Chimera~\cite{PettersenHuangCouchGreenblattMengFerrin2004}.
		The HRSI and HRSS reconstructions use different color maps.
		All markings are scaled by $10^{-3}$.%\vspace*{-.2in}
	}
\end{figure}
Fig.~\ref{fig:stdv_d} (b) and (c) show the standard deviation $s_\rho(\vx)$ computed by HRSI and HRSS, respectively. In Fig.~\ref{fig:rhobar_d}, $\bar\rho(\vx)$ computed by HRSI and the two heterogeneous solutions  $s_\rho(\vx)$ computed by HRSI and HRSS are jointly visualized by computing a 3-D surface of constant value of $\bar\rho(\vx)$
and coloring the surface by $s_\rho(\vx)$. The cubes of $\bar\rho$ and $s_\rho$ in Fig.~\ref{fig:stdv_d}--\ref{fig:rhobar_d} have the same orientation as shown in Fig.~\ref{fig:stdv_d} (a), in which the x-z plane contains the two fold symmetry of the icosahedron. We call Fig.~\ref{fig:stdv_d} (a) as the two fold axes.

\begin{figure}[t]
	\begin{center}
		\begin{tabular}{c}
			\includegraphics[height=26mm]{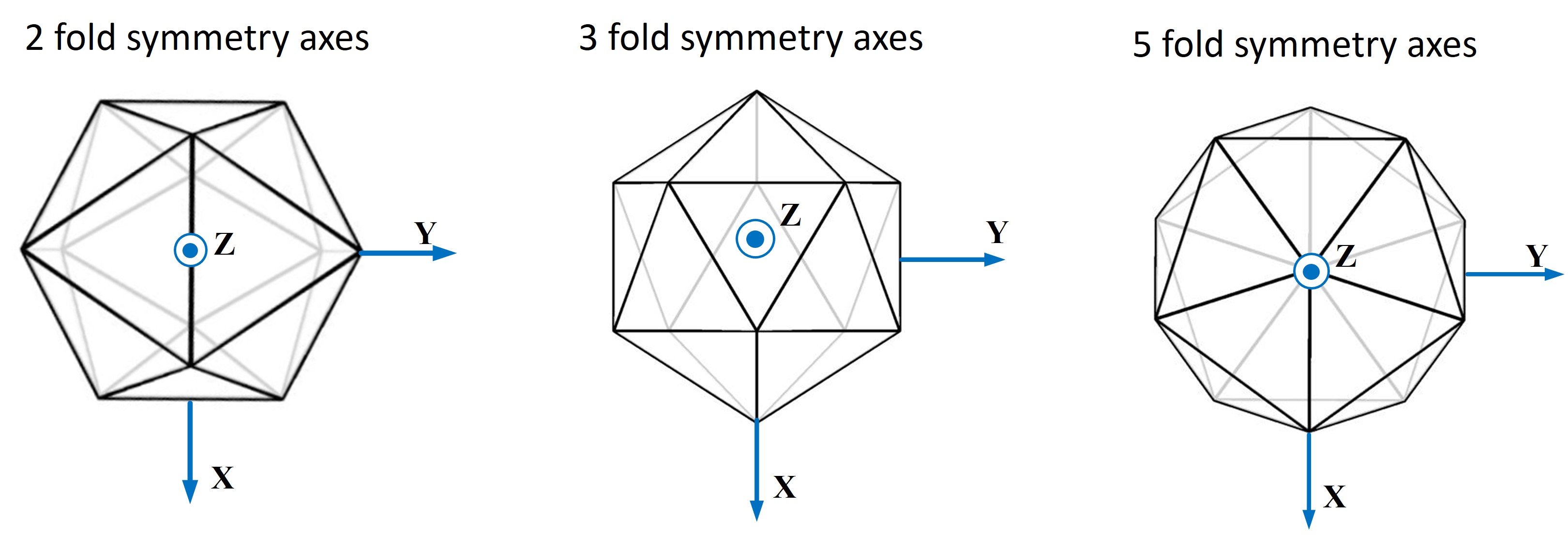}\\
			
			\includegraphics[height=56mm]{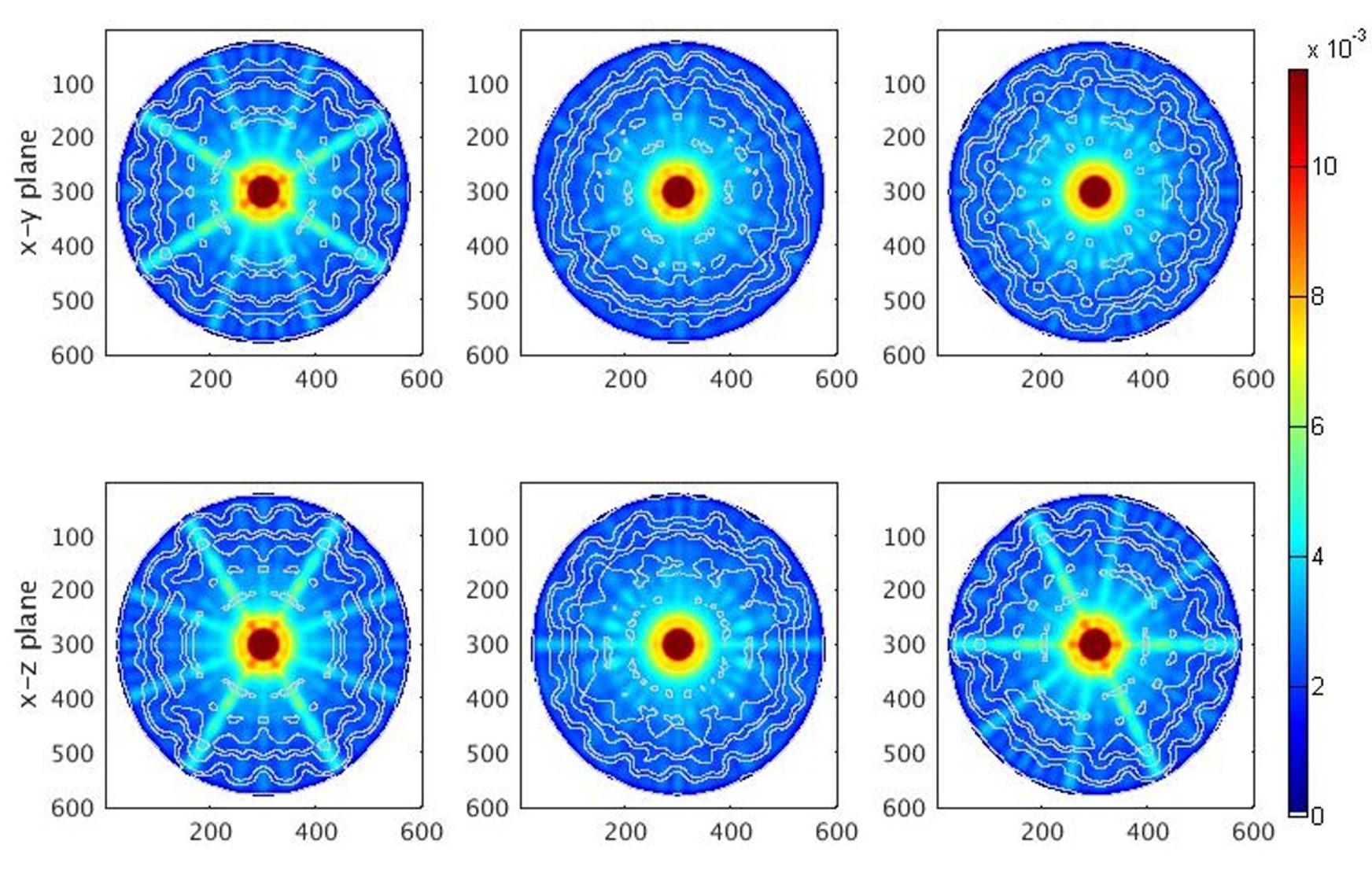}\\
			(a) symmetrical individuals (HRSI)\\
			
			\includegraphics[height=56mm]{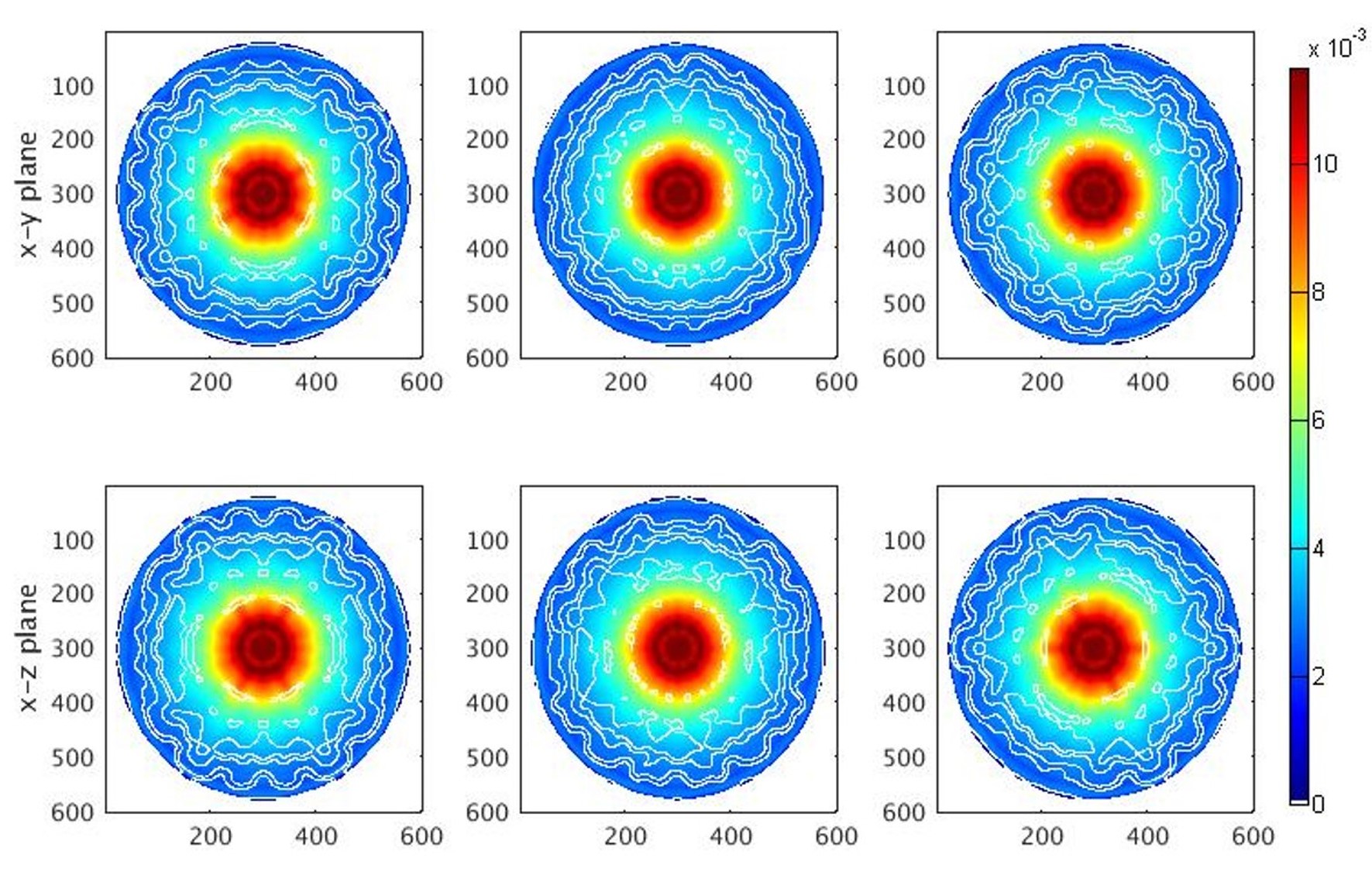}\\			
			(b) symmetrical statistics (HRSS)
		\end{tabular}
	\end{center}
	%\vspace*{-.15in}
	\caption{
		\label{fig:stdv}
		Cross section of the standard deviation function $s_{\rho}(\vx)$ for HK97 Prohead
		I$\textsuperscript{+pro}$ using both HRSI and HRSS and displayed with a common color map. The white lines depict the contour levels of the virus particles. The geometrical orientation of the cubes are shown in the top row. (a) shows the reconstruction results based on symmetrical individuals, i.e., HRSI, whereas (b) shows the reconstruction results based on symmetrical statistics, i.e., HRSS. For each orientation, cross sections in x-y and x-z planes are displayed. The center of the $s_\rho$ cube has the coordinate (300, 300, 300). 		
	}
\end{figure}

Fig.~\ref{fig:stdv} shows six different cross sections of the standard deviation function $s_\rho(\vx)$ in three different axes the two fold, the three fold and the five fold symmetry axes. The two fold symmetry, the three fold symmetry and the five fold symmetry of the icosahedron are sitting along the Z axis, as shown in the top row of Fig.~\ref{fig:stdv}, where the center of the $s_\rho$ cube has the coordinate (300, 300, 300). 

\par
%There are two primary observations.
As shown in Fig.~\ref{fig:stdv_d} (b) and Fig.~\ref{fig:rhobar_d} (a), the existing algorithm HRSI estimates a standard deviation function that is organized in radially-directed rays, which has no biological explanation. Specifically, HRSI introduced artifical spikes laying on the symmetry axes of the icosahedron. In fact, this artificial pattern in standard deviation or variance analysis of spherical virus particles has been recognized as a well-known problem of all existing reconstruction algorithms~\cite[p.~173]{ludtke2016chapter}, including the standard algorithms EMAN2 \cite{tang2007eman2} and SPIDER~\cite{shaikh2008spider}. 

In contrast, HRSS gives a standard deviation function that is organized in \textit{annular} structures as shown in Fig.~\ref{fig:stdv_d} (c) and more obviously in Fig.~\ref{fig:rhobar_d} (b). Such an annular structure obtained by HRSS matches the physical organization of the particle, for instances, the outer protein shell and the inner core of nucleic acid. This makes more biological sense than the estimates of HRSI and other reconstruction algorithms. Such a distinction can be further seen in Fig.~\ref{fig:stdv}, in which Fig.~\ref{fig:stdv} (a) has light blue radial lines, whereas the Fig.~\ref{fig:stdv} (b) shows the detailed annular structure of the virus. In the previous analyses with the use of HRSI, features like high values located along radially directed lines averaged out leaving interpretable information, but as we try to increase the spatial detail of our interpretation, such behavior is difficult to understand.
\section{Conclusion}\label{sec:conc}
In this paper, we proposed a novel approach HRSS for 3-D reconstruction of nanoscale virus particles from noisy 2-D projection images. Realistic assumption on the heterogeneous virus particles requires sophisticated tools.
HRSS incorprates the complete basis functions to realize
the sophisticated situation of symmetrical statistics, which eliminates the long-standing distortions in heterogeneity calculations associated with symmetry axes, and provides estimates that make more biological sense than the previous estimates. This implies that the variability in the particles is such that individual particles lack symmetry and only the first and second order {\it statistics} obey the symmetry.
This is natual for such large objects as viruses ($10^2-10^3${$\AA$}, 10 Megadalton) in the absence of the additional geometric constraints that occur in the crystal of an x-ray crystallography experiments.

\section{Acknowledgments}
%\vspace*{-.1in}
We are grateful to Professors John E. Johnson (The Scripps Research
Institute) and David Veesler (University of Washington) for the HK97 data
and helpful discussions.

\appendices
\section{Derivation of Eq.~\ref{eq:Fortho}}
The orthonormal condition of angular basis functions $I_{p;l,n}$ (Property 2 in Section~\ref{sec:BasisFunction}) explicitly is~\cite{XuBasisFunctions2016}
\begin{equation}\label{eq:Iortho}
\int_{\theta=0}^{\pi}\int_{\phi=0}^{2\pi}I_{p;l,n}^T(\vx/x)I_{p';l',n'}(\vx/x)\dd\theta\dd\phi=I_{d_p}\delta_{p,\pp}\delta_{l,l'}\delta_{n,n'},\end{equation}
where $\vx/x$ is shorthand for $(\theta,\phi)$.
By Eq.~\ref{eq:Iortho} and Eq.~\ref{eq:radialortho} (the orthonomality of radial basis functions), the LHS of~\ref{eq:Fortho} can be expressed as
{\scriptsize
\begin{equation}
\begin{split}
&\int_{\vx}F_{p;\zeta}(\vx) F_{p';\zeta'}^T(\vx)\dd\vx\\
&=\int_{\vx}\left(I_{p;l,n}(\vx/\Radius)h_{l,q}(\Radius)\right)
\left(I_{p';l',n'}(\vx/\Radius)h_{l',q'}(\Radius)\right)^T
\dd\vx\\
&=\int_{\theta=0}^{\pi}\int_{\phi=0}^{2\pi}\int_{\Radius=0}^{\infty}\left(I_{p;l,n}(\vx/\Radius)h_{l,q}(\Radius)\right)
\left(I_{p';l',n'}(\vx/\Radius)h_{l',q'}(\Radius)\right)^T
\dd\Radius\dd\phi\dd\theta\\
&=\int_{\theta=0}^{\pi}I_{p;l,n}(\vx/\Radius)\left(\int_{\phi=0}^{2\pi}h_{l,q}(\Radius)h_{l',q'}^T(\Radius)\dd\Radius\right)
I_{p';l',n'}^T(\vx/\Radius)
\dd\phi\dd\theta\\
&=\left(\delta_{l,l'}\delta_{q,q'}\right)\int_{\theta=0}^{\pi}I_{p;l,n}(\vx/\Radius)
I_{p';l,n'}^T(\vx/\Radius)
\dd\phi\dd\theta\\
&=I_{d_p}\delta_{p,\pp}\delta_{l,l'}\delta_{n,n'}\delta_{q,q'}\\
&=I_{d_p}\delta_{p,\pp}\delta_{\zeta,\zeta'}
\end{split}
\end{equation}
}

\section*{Acknowledgment}
We are grateful to Professors John E. Johnson (The Scripps Research
Institute) and David Veesler (University of Washington) for the HK97 data and helpful discussions.

\ifCLASSOPTIONcaptionsoff
  \newpage
\fi
\bibliographystyle{IEEEtran}
\bibliography{referencesnB,mybibfileFINAL}\vspace*{-1cm}

\begin{thebibliography}{10}
\providecommand{\url}[1]{#1}
\csname url@samestyle\endcsname
\providecommand{\newblock}{\relax}
\providecommand{\bibinfo}[2]{#2}
\providecommand{\BIBentrySTDinterwordspacing}{\spaceskip=0pt\relax}
\providecommand{\BIBentryALTinterwordstretchfactor}{4}
\providecommand{\BIBentryALTinterwordspacing}{\spaceskip=\fontdimen2\font plus
\BIBentryALTinterwordstretchfactor\fontdimen3\font minus
  \fontdimen4\font\relax}
\providecommand{\BIBforeignlanguage}[2]{{%
\expandafter\ifx\csname l@#1\endcsname\relax
\typeout{** WARNING: IEEEtran.bst: No hyphenation pattern has been}%
\typeout{** loaded for the language `#1'. Using the pattern for}%
\typeout{** the default language instead.}%
\else
\language=\csname l@#1\endcsname
\fi
#2}}
\providecommand{\BIBdecl}{\relax}
\BIBdecl

\bibitem{JensenMethodsEnzymologyABC2010}
G.~J. Jensen, Ed., \emph{Cryo-EM, Parts A--C}, ser. Methods in
  Enzymology.\hskip 1em plus 0.5em minus 0.4em\relax Elsevier Inc., 2010, vol.
  481--483.

\bibitem{QiuWangMatsuiDomitrovicYiliZhengDoerschukJohnsonJSB2012}
Q.~Wang, T.~Matsui, T.~Domitrovic, Y.~Zheng, P.~C. Doerschuk, and J.~E.
  Johnson, ``Dynamics in cryo {EM} reconstructions visualized with
  maximum-likelihood derived variance maps,'' \emph{J. Struct. Biol.}, vol.
  181, no.~3, pp. 195--206, Mar. 2013.

\bibitem{YiliZhengQiuWangDoerschukJOSA2012}
Y.~Zheng, Q.~Wang, and P.~C. Doerschuk, ``3-{D} reconstruction of the
  statistics of heterogeneous objects from a collection of one projection image
  of each object,'' \emph{J. Opt. Soc. Am. A}, vol.~29, no.~6, pp. 959--970,
  Jun. 2012.

\bibitem{baron1996epidemiology}
S.~Baron, \emph{Epidemiology--Medical Microbiology}.\hskip 1em plus 0.5em minus
  0.4em\relax University of Texas Medical Branch at Galveston, 1996.

\bibitem{DoerschukJohnsonIT2000}
P.~C. Doerschuk and J.~E. Johnson, ``{\em Ab initio} reconstruction and
  experimental design for cryo electron microscopy,'' \emph{IEEE Transactions
  on Information Theory}, vol.~46, no.~5, pp. 1714--1729, Aug. 2000,
  {\url{http://dx.doi.org/10.1109/18.857786}}.

\bibitem{YinZhengDoerschukNatarajanJohnsonJSB2003}
Z.~Yin, Y.~Zheng, P.~C. Doerschuk, P.~Natarajan, and J.~E. Johnson, ``A
  statistical approach to computer processing of cryo electron microscope
  images: {V}irion classification and 3-{D} reconstruction,'' \emph{J. Struct.
  Biol.}, vol. 144, no. 1/2, pp. 24--50, 2003,
  {\url{http://www.ncbi.nlm.nih.gov/pubmed/14643207}}.

\bibitem{ScheresGaoValleHermanEggermontFrankCarazoNatureMethods2007}
S.~H.~W. Scheres, H.~Gao, M.~Valle, G.~T. Herman, P.~P.~B. Eggermont, J.~Frank,
  and J.-M. Carazo, ``Disentangling conformational states of macromolecules in
  3{D}-{EM} through likelihood optimization,'' \emph{Nature Methods}, vol.~4,
  no.~1, pp. 27--29, Jan. 2007.

\bibitem{JLeeDoerschukJohnsonIP2007}
J.~Lee, P.~C. Doerschuk, and J.~E. Johnson, ``Exact reduced-complexity maximum
  likelihood reconstruction of multiple 3-{D} objects from unlabeled unoriented
  2-{D} projections and electron microscopy of viruses,'' \emph{IEEE
  Transactions on Image Processing}, vol.~16, no.~11, pp. 2865--2878, Nov.
  2007, {\url{http://www.ncbi.nlm.nih.gov/pubmed/18092587}}.

\bibitem{RednerWalker1984}
R.~A. Redner and H.~F. Walker, ``Mixture densities, maximum likelihood and the
  {EM} algorithm,'' \emph{SIAM Review}, vol.~26, no.~2, pp. 195--239, Apr.
  1984.

\bibitem{XuBasisFunctions2016}
N.~Xu and P.~C. Doerschuk, ``Computation of real basis functions for the 3-d
  rotational polyhedral groups $t$, $o$, and $i$,'' \emph{Submitted to SIAM
  Journal of Mathematical Analysis}, 2016.

\bibitem{OlverLozierBoisvertClark2010}
\BIBentryALTinterwordspacing
F.~W.~J. Olver, D.~W. Lozier, R.~F. Boisvert, and C.~W. Clark, Eds., \emph{NIST
  Handbook of Mathematical Functions}.\hskip 1em plus 0.5em minus 0.4em\relax
  Cambridge, UK: Cambridge University Press, 2010. [Online]. Available:
  \url{http://dlmf.nist.gov}
\BIBentrySTDinterwordspacing

\bibitem{Cornwell1984}
J.~F. Cornwell, \emph{Group Theory in Physics}.\hskip 1em plus 0.5em minus
  0.4em\relax London: Academic Press, 1984, vol.~1.

\bibitem{ZhengDoerschukIP1996}
Y.~Zheng and P.~C. Doerschuk, ``3{D} image reconstruction from averaged
  {F}ourier transform magnitude by parameter estimation,'' \emph{IEEE
  Transactions on Image Processing}, vol.~7, no.~11, pp. 1561--1570, Nov. 1998.

\bibitem{TangKearneyQiuWangDoerschukBakerJohnsonJMolRecog2014}
J.~Tang, B.~M. Kearney, Q.~Wang, P.~C. Doerschuk, T.~S. Baker, and J.~E.
  Johnson, ``Dynamic and geometric analyses of {{\em Nudaurelia capensis}}
  $\omega$ virus maturation reveal the energy landscape of particle
  transitions,'' \emph{J. Molecular Recognition}, vol.~27, no.~4, pp. 230--237,
  10 February 2014, {\url{http://www.ncbi.nlm.nih.gov/pubmed/24591180}}.

\bibitem{vanHeelUltramicroscopy1987}
M.~van Heel, ``Similarity measures between images,'' \emph{Ultramicroscopy},
  vol.~21, pp. 95--100, 1987.

\bibitem{HarauzvanHeelOptik1986}
G.~Harauz and M.~van Heel, ``Exact filters for general geometry three
  dimensional reconstruction,'' \emph{Optik}, vol.~73, no.~4, pp. 146--156,
  1986.

\bibitem{BakerOlsonFullerMicrobiolMolBiolRev1999}
T.~S. Baker, N.~H. Olson, and S.~D. Fuller, ``Adding the third dimension to
  virus life cycles: Three-dimensional reconstruction of icosahedral viruses
  from cryo-electron micrographs,'' \emph{Microbiology and Molecular Biology
  Reviews}, vol.~63, no.~4, pp. 862--922, Dec. 1999.

\bibitem{PettersenHuangCouchGreenblattMengFerrin2004}
E.~F. Pettersen, T.~D. Goddard, C.~C. Huang, G.~S. Couch, D.~M. Greenblatt,
  E.~C. Meng, and T.~E. Ferrin, ``{UCSF} {C}himera---{A} visualization system
  for exploratory research and analysis,'' \emph{J. Comput. Chem.}, vol.~25,
  no.~13, pp. 1605--1612, 2004.

\bibitem{ludtke2016chapter}
S.~Ludtke, ``Chapter seven-single-particle refinement and variability analysis
  in eman2. 1,'' \emph{Methods in Enzymology}, vol. 579, pp. 159--189, 2016.

\bibitem{tang2007eman2}
G.~Tang, L.~Peng, P.~R. Baldwin, D.~S. Mann, W.~Jiang, I.~Rees, and S.~J.
  Ludtke, ``Eman2: an extensible image processing suite for electron
  microscopy,'' \emph{Journal of structural biology}, vol. 157, no.~1, pp.
  38--46, 2007.

\bibitem{shaikh2008spider}
T.~R. Shaikh, H.~Gao, W.~T. Baxter, F.~J. Asturias, N.~Boisset, A.~Leith, and
  J.~Frank, ``Spider image processing for single-particle reconstruction of
  biological macromolecules from electron micrographs,'' \emph{Nature
  protocols}, vol.~3, no.~12, pp. 1941--1974, 2008.

\end{thebibliography}
\begin{IEEEbiography}[{\includegraphics[width=1in,height=1.25in,clip,keepaspectratio]{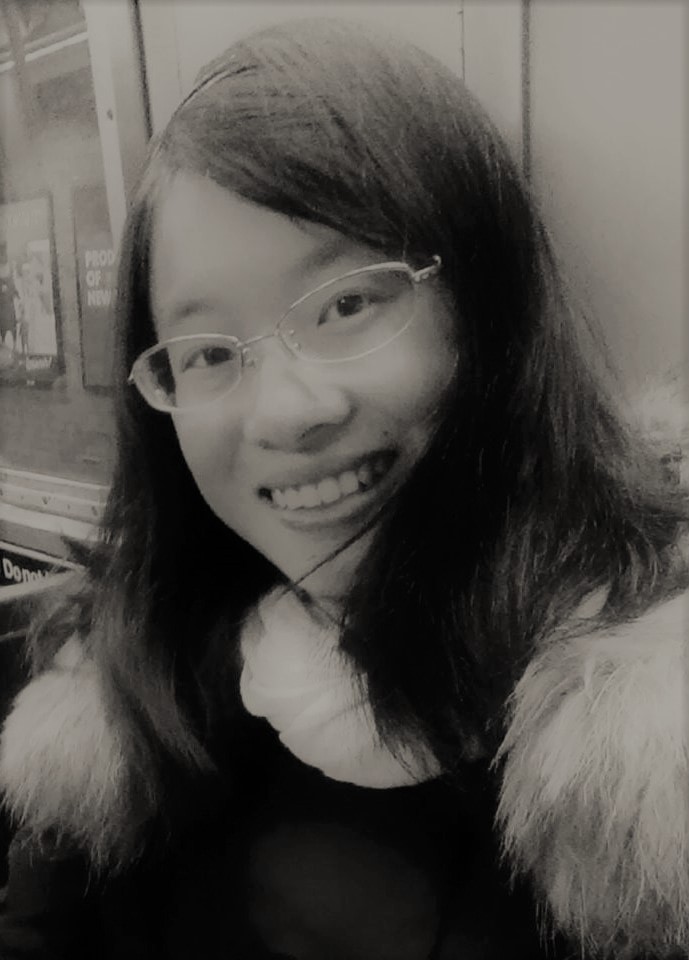}}]%
{Nan Xu} is pursuing her Ph.D. degree in School of Electrical and Computer Engineering with a minor in Applied Mathematics and a minor in Cognitive Neuroscience at Cornell University. Nan received her Master of Science degree in May 2015. Before joining Cornell, she received double Bachelor's degrees in Electrical and Computer Engineering (BS) and Mathematics (BA) with a minor in Music at the University of Rochester in Rochester NY. Nan's current research interests are in statistical modeling and inference in biological data. She is investigating three primary applications: brain network estimation, realistic fMRI data simulation, and reconstruction of the statistical characteristics of ensembles of heterogeneous virus particles.
\end{IEEEbiography}\vspace*{-2cm}

\begin{IEEEbiography}[{\includegraphics[width=1in,height=1.25in,clip,keepaspectratio]{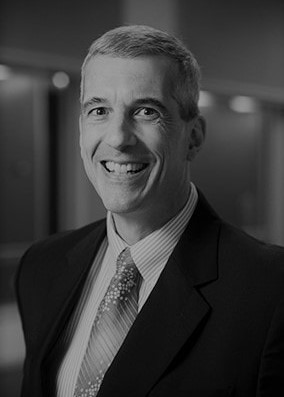}}]%
{Peter C Doerschuk} 
is a professor in the School of Electrical and Computer Engineering and Nancy E. and Peter C. Meing School of Biomedical Engineering at Cornell University. His research concerns biological and medical systems from the view point of computational nonlinear stochastic systems. In particular, he has contributed to computational inverse problems for biophysics, statistical image processing, and biomedical and speech signal processing. In each of these areas, important goals of his work have been to incorporate accurate physical models while at the same time developing computationally practical algorithms implemented in high-performance software systems.
\par
Before joining Cornell University in July 2006, Peter Doerschuk was on the faculty of Purdue University in Electrical and Computer Engineering and Biomedical Engineering. He received BS, MS, and Ph.D. degrees in Electrical Engineering from MIT and an M.D. degree from Harvard Medical School. After post-graduate training at Brigham and Womens' Hospital he held a post-doctoral appointment at the Laboratory for Information and Decision Systems (MIT) before joining Purdue.
\end{IEEEbiography}
\end{document}